\documentclass[twocolumn,showpacs,preprintnumbers,amsmath,amssymb]{revtex4}

\usepackage{graphicx}
\usepackage{dcolumn}
\usepackage{bm}

\begin{document}


\title{ Superconducting vortex profile from fixed point measurements\\ The Lazy Fisherman tunnelling microscopy  method}

\author{A. Kohen*, T. Cren*, Th. Proslier*, Y. Noat*, F. Giubileo, F. Bobba,  A. M. Cucolo, W. Sacks* and D. Roditchev*}
\affiliation{%
Physics Dept. and INFM-SUPERMAT Lab., University of Salerno, via
S. Allende, 84081 Baronissi (SA), Italy
\\
*Institut des Nanosciences de Paris, Universit\'es Paris 6 et 7,
C.N.R.S. (UMR 75\ 88),  75015 Paris, France
}%

\author{N. Zhigadlo, S.M. Kazakov, J. Karpinski}
\affiliation{ Solid State Physics Laboratory, ETH Zurich, CH-8093
Zurich, Switzerland
}%

\date{\today}

\begin{abstract}
We introduce a mode of operation for studying  the vortex phase in
superconductors using scanning tunnelling microscopy (STM). While
in the conventional STM method, the tip is scanned over a sample
in which a fixed vortex pattern is prepared, in our "Lazy
Fisherman" method the STM tip is kept fixed at a selected location
while the vortices are being moved by varying the applied magnetic
field. By continuously acquiring the local tunnelling conductance
spectra, dI/dV(V), we detect the changes in the local density of
states under the tip due to the vortex motion. With no need for
scanning, the method permits one to extend the study of vortices
to samples in which scanning is difficult or even impossible due
to surface nonuniformity and allows one to study vortex dynamics.
Using a statistical analysis of the spectra, we reconstruct the
single vortex zero bias conductance profile. We apply the method
to the c-axis face of an MgB$_2$ single crystal sample and obtain
a vortex profile with a coherence length, $\xi$ of 57$\pm$2 nm.
\end{abstract}

\maketitle

In type II superconductors the magnetic field penetrates the
superconducting (SC) material, above a certain field (Hc$_1$), in
the form of quantized flux bundles, known as vortices, each vortex
carrying a flux quantum of $\phi_0$=hc/e \cite{Abrikosov}. The
vortices arrange themselves in a lattice and the SC order
parameter (OP), $\Delta(\textbf{r})$, becomes spatially
inhomogeneous, having a zero value in the center of each vortex
and rising to its maximum value in between the vortices. Several
experimental methods were developed for studying the vortex
lattice. Most methods exploit the spatial variations in the value
of the local magnetic field and are therefore sensitive to changes
occurring over a length scale set by the magnetic field
penetration length, $\lambda$ . These include: Bitter decoration
\cite{Bitter}, magneto-optical microscopy \cite{Magnopt}, local
hall microscopy \cite{Hall}, electron interference microscopy
\cite{Lorentz} and squid microscopy \cite {SQUID}. A different
approach was developed using a scanning tunnelling microscope
(STM)\cite{Hess}. The STM is used to map the electronic density of
states (DOS) as a function of position by scanning its tip over
the sample. This enables one to view the vortex lattice based on
the observed changes in the value of the Zero Bias Conductance
(ZBC) which manifest the changes in value of $\Delta(\textbf{r})$.
Therefore the method is sensitive to changes occurring on a length
scale set by the SC coherence length, $\xi$. This, together with
the STM's ability to obtain atomic resolution in the topographic
mode, results in images of extremely high spatial resolution. The
method had been successfully used in a number of materials such as
NbSe$_{2}$ \cite{Hess}, BSCCO \cite{Pan},
YBCO\cite{Maggio-Aprile}, LuNi$_{2}$B$_{2}$C\cite{Dewilde} and
MgB$_2$\cite{Eskildsen}. A main drawback of this method is that
its use is limited to highly flat samples. Moreover the relatively
large time needed for obtaining a single image (several minutes
for a picture obtained at a single voltage up to hours for the
full spectra) limits its use in the study of vortex dynamics.

\begin{figure}
\includegraphics[angle=-90,width=8cm]{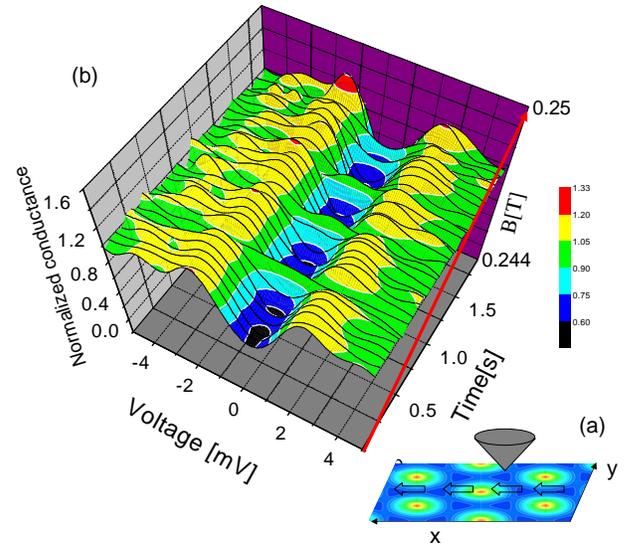}
\caption{\label{fig:epsart1} a) A schematic drawing of the
experimental method showing the STM tip (conic shape) fixed at a
chosen position and the vortex motion (marked by arrows). b) A
color coded 3D plot showing 30 normalized conductance spectra as a
function of voltage and time exhibiting the variations in the
local DOS under the tip induced by vortex motion. The magnetic
field time dependence inducing the vortex motion is given by the
thick line }
\end{figure}

In this letter we show how to extend the use of STM to samples in
which scanning is difficult or even impossible by using a
different mode of operation which we name the lazy fisherman
method (LFM) \cite{note}. In LFM the difficulties involved in
scanning are avoided by keeping the STM tip at a fixed position in
which at zero field a clear SC spectra is observed. The different
points within the vortex lattice are accessed by creating a vortex
motion, induced by slowly changing the value of the applied field,
as illustrated in Fig.1a. We show how by statistically analyzing
the data measured at a \textbf{single fixed} point one can
reconstruct the vortex profile and thus extract the SC coherence
length. We apply the LFM to a c-axis oriented surface of a single
crystal MgB$_{2}$ sample. We find that the measured data
represents well a statistical ensemble sampled uniformly over the
vortex unit cell. We reconstruct the vortex profile in MgB$_2$ and
find the SC coherence length to be $\xi$=57$\pm$2 nm in close
agreement with the value of 51 nm obtained by Eskildsen et
al.\cite{Eskildsen} using the conventional STM scanning method.

Single crystals of MgB$_2$ were grown by a high pressure method
described in Ref. \cite{Karpinski}. The experiments were carried
out on a low temperature Omicron STM. The tunneling junctions were
achieved by approaching mechanically cut Pt/Ir tips to the c-axis
oriented surface of the crystal. Due to the peculiar electronic
structure of the material, c-axis tunnelling allows to probe
mainly the quasiparticle density of states in the $\pi$-band
\cite{Brinkman,Eskildsen}. The superconducting critical
temperature was determined locally by measuring the evolution of
the tunneling conductance spectra as a function of temperature,
and was found to be 38.5 K. The field dependence of the tunneling
conductance spectra $dI(V)/dV$ was obtained by fixing the STM tip
in a selected location and continuously measuring local $I-V$
tunnelling curves while slowly sweeping the magnetic field, the
field being parallel to the c-axis. At low temperature (T = 6.6
K), the field was increased from zero up to the maximum value of
2.3 T at a rate of  0.15 T/min. The acquisition time for a single
spectrum (-20 mV, 20 mV) was 60 msec. However the voltage range in
which the DOS is effected by the vortex, $\pm\Delta\approx\pm$ 2
mV), is scanned in only 6 msec. During such a short time the field
may be considered constant for an individual spectra, as $\Delta B
\simeq 10^{-5}$ T, thus $\Delta B / B << 1$. As our measured
spectra are symmetrical with respect to zero bias we conclude
that, during the time required to obtain a single spectrum, the
vortex motion is also negligible. However subsequently measured
spectra do differ as can be seen in Fig.1b which shows an example
of 30 consecutively acquired spectra obtained in the field range
of 0.244 T to 0.250 T. One clearly sees the non monotonic changes
in the local DOS under the tip, due to the vortex motion. The
spectra range from a clearly gapped form showing shoulders at the
gap edges (reflecting the DOS far from the vortex core) to a
completely flat form (reflecting the DOS at the vortex core).

\begin{figure}
\includegraphics[angle=0, width=8cm]{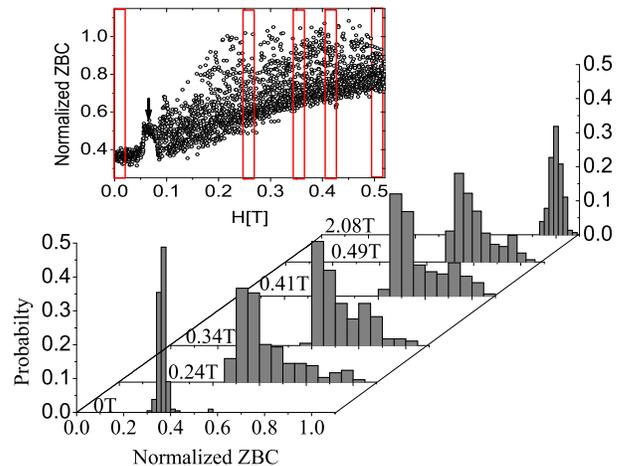}

\caption{\label{fig:epsart}Field evolution of the Normalized ZBC
histograms: 0 T, 0.24 T, 0.34 T, 0.41 T, 0.49 T and 2.08 T.
Histograms obtained at intermediate field values (0.24 T - 0.49 T)
show an asymmetric distribution peaked at or slightly above the
minimum, reflecting the shape of the vortex profile. Inset shows
the ZBC values obtained for 0$ T < $H$ < $0.5 T. Rectangles mark
the ranges used for obtaining the histograms shown in the main
figure}
\end{figure}

A commonly used parameter in conventional STM vortex mapping is
the value of the normalized ZBC, $\sigma\equiv$
$[dI/dV(V=0)]/[dI/dV(V>>\Delta)]$. In the normal state this value
is equal to 1 while in the SC state it is strongly reduced due to
an opening of a gap in the DOS around the Fermi level. Thus, one
obtains the vortex image by mapping the value of the ZBC as a
function of position. The ZBC obtains its maximum value at the
vortex core, gradually decreases as a function of the distance
from the vortex core and reaches its minimum value at the point of
equidistance from the vortices (see for example \cite{Eskildsen}).
Measuring the value of the ZBC as a function of the distance from
the vortex core allows one to obtain the exact vortex profile and
thus to estimate the SC coherence length, $\xi$.

As in our method one can not directly determine the distance
between the vortex core and the STM's tip we use a statistical
method in order to study the distribution of the ZBC values and to
reconstruct the vortex profile. As a first step we have created
histograms of the ZBC values acquired over 10 sec intervals. The
time interval was chosen to be on one hand short enough to assure
that the inter-vortex distance does not significantly change
($\Delta d/d<6\% $) and on the other hand  to allow enough data
points for statistical analysis.

The inset of Fig.2 shows the ZBC values as a function of field for
0$ < $H$ < $0.5 T. The field in which the vortex start to pass
under the tip is clearly seen as a sharp jump in the values of the
ZBC (see arrow in the inset). As this value is always larger than
Hc$_{1}$ due to surface barrier effects we conclude that
Hc$_{1||c}\lesssim0.12\pm0.03 T$ in agreement with the value of
0.1 T found by Lyard et al.\cite{lyard}  by magnetization
measurements. The resulting histograms at fields of 0 T, 0.24 T,
0.34 T, 0.41 T, 0.49 T and 2.08 T are shown in Fig.2. Each
histogram represents roughly 150 ZBC values obtained in a field
range of 0.03 T. The zero field histogram is narrow, symmetric and
as expected shows no sign of spread in the ZBC values due to
vortices. At higher fields (H$ > $0.1 T) the histograms become
highly asymmetric. The probability is peaked near the minimum
conductance value and decays considerably towards the maximum
value. Looking at the set of histograms one can see an increase of
the minimum value as the field increases, reflecting the
increasing overlap between the vortices. In the case of MgB$_{2}$
the exact field dependence of this increase is rather complicated
due to coupling between the two bands. We will discuss the
properties related to the unique two band nature of MgB$_{2}$ in a
separate publication \cite{kohen2}. Finally at high fields (H$> $2
T) the histogram becomes symmetric again with the modulation due
to the vortices being so small that it is almost entirely washed
out by the experimental noise. As no gap can be detected in any of
our spectra for H$>$2.3T we deduce H$_{c2||c}(T=6.6 K)\gtrsim$ 2.3
T, a value comparable to that measured by Lyard et al.\cite{lyard}
of 2.6 T.

\begin{figure}
\includegraphics[width=5.7cm]{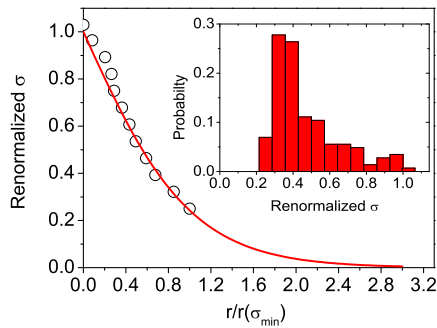}
\caption{vortex profile, $\sigma$(r) (circles) with a theoretical
dependence given by 1-tanh(r/$\xi$)(line). Best fit is given by
$\xi$=57$\pm2$nm. Inset shows the Renormalized ZBC values
histogram used to calculate the vortex profile }
\end{figure}

Using the ZBC histograms we are able to reconstruct, with no free
parameters, the vortex profile, ZBC(r)(=$\sigma$(r)), where r is
the distance from the vortex core. Fig.3 shows the results
calculated from the experimental histogram obtained in the field
range of 0.24-0.27 T (see inset of Fig.3). The abscissa which is
the distance from the vortex core is calculated directly from the
histogram using $r(\sigma_{n})/r(\sigma_{min})$=$\sqrt{\sum_{i=n+1
\rightarrow N}P(\sigma_{i},\sigma_{i}+\delta\sigma)}$, where
$\delta\sigma$ is the bin size. This expression is exact for a
vortex profile of cylindrical symmetry. In such a case the
probability of obtaining a ZBC value between $\sigma_1$ and
$\sigma_2$ which is general given by
P($\sigma_{1}$,$\sigma_{2}$)$\propto\int{d \theta
(r^{2}(\sigma_{2},\theta)-r^{2}(\sigma_{1},\theta))}$ reduces to
P($\sigma_{1}$,$\sigma_{2}$)$\propto
(r^{2}(\sigma_{2})-r^{2}(\sigma_{1})$ . This approximation is
valid as long as the inter-vortex distance is large in comparison
to the coherence length ($\xi<$d) and that the material is in the
dirty limit ($\xi>\ell$), which is the case in
MgB$_{2}$\cite{Eskildsen}. The ZBC is presented on a renormalized
scale $(\sigma(H)-\sigma(H=0))/(1-\sigma(H=0))$. This allows one
to separate the contribution of DOS changes to the ZBC from that
of finite temperature effects and inelastic tunnelling process.
The solid line in the figure is a best fit curve to the data
obtained using $\sigma$(r)=$1-tanh(r/\xi)$. The best fit is
obtained for a value of $\xi/r(\sigma_{min})$=0.99$\pm$0.03. As
the minimum ZBC value is obtained (for a triangular lattice) at
r($\sigma_{min}$) = d/$\sqrt{3}$=59 nm (for H = 0.24 T), where d
is the inter vortex distance given by d $\simeq$ 50 nm/$\sqrt{H}$,
this results in $\xi$ = 57 $\pm$2 nm.

To conclude we have presented a method which allows to extend the
use of STM for vortex study to cases where scanning is difficult
or impossible and to cases showing fast vortex dynamics. Using the
method one can measure several basic parameters describing the SC
vortex state such as the field of first vortex penetration, the SC
coherence length $\xi$ and the upper critical field Hc$_{2}$. At
the same time our method maintains the scanning capabilities of
the STM. Thus, it allows repeating the measurement at different
fixed points in the sample and relating the observed changes in
vortex behavior to the position in the sample. This could be of
interest in the vicinity of impurities, pinning centers and grain
boundaries. Additionally the LFM is suitable in all type of
samples (rough and flat) for studying faster vortex dynamics (in
comparison to conventional scanning) being limited only by the
time needed to measure a \textbf{single} spectrum, $\sim$1ms. A
variant of the method can be used for studying faster vortex
dynamics by measuring the tunnelling current (with open feedback
loop) at a voltage of $eV\lesssim\Delta$. As vortex motion is the
source of dissipation in the SC state, LFM being a local fast tool
to study the SC parameters related to the vortex lattice and the
vortex dynamics could of great use in the development of SC
applications.

The authors thank Marco Aprili and Vincent Jeudi for useful
discussions. This work has been supported by Italian MIUR project
"$\it{Rientro \, dei \, Cervelli}$" and by the French University
Paris 6 PPF project.



\thebibliography{apssamp}
\bibitem{Abrikosov} A. A. Abrikosov Soviet Physics JETP \textbf{5}, 1174 (1957)
\bibitem{Bitter} F. Bitter  Phys. Rev. B \textbf{38}, 1903 (1931)
\bibitem{Magnopt} P. B. Alers Phys Rev \textbf{105}, 104 (1957)
\bibitem{Hall} P. Leiderer, P. Brüll, T. Klumpp and B.Stritzker Physica B \textbf{165-166}, 1387 (1990)

\bibitem{Lorentz} T. Matsuda, S. Hasegawa, M. Igarashi, T. Kobayashi, M. Naito, H. Kajiyama, J. Endo, N. Osakabe, A. Tonomura
and R. Aoki  Phys Rev Lett \textbf{62}, 2519 (1989)

\bibitem{SQUID} J.R. Kirtely, M.B. Ketchen, K. G. Stawiasz, J.Z. Sun, W.J. Gallagher, S.H. Blanton and S.j. Wind  Appl. Phys. Lett. \textbf{66} 1138 (1995)

\bibitem{Hess} H. F. Hess, R. B. Robinson, R. C. Dynes, J. M. Valles, Jr. and J. V. Waszczak  Phys. Rev. Lett. \textbf{62}, 214 (1989)


\bibitem{Pan}S. H. Pan, E. W. Hudson, A. K. Gupta, K.-W. Ng, H. Eisaki, S. Uchida, and J. C. Davis  Phys. Rev. Lett. \textbf{85}, 1536 (2000)
\bibitem{Maggio-Aprile} I. Maggio-Aprile, Ch. Renner, A. Erb, E. Walker, and Ø. Fischer  Phys. Rev. Lett. \textbf{75}, 2754 (1995)
\bibitem{Troyanovski} GJC van Baarle, AM Troianovski, T. Nishizaki, PH Kes, and J. Aarts  Appl. Phys. Lett. \textbf{82}, 1081 (2003)
\bibitem{Dewilde} Y. De Wilde, M. Iavarone, U. Welp, V. Metlushko, A. E. Koshelev, I. Aranson, and G. W. Crabtree Phys. Rev. Lett.
\textbf{78}, 4273 (1997)
\bibitem{Eskildsen} M. R. Eskildsen, M. Kugler, S. Tanaka, J. Jun, S. M. Kazakov, J. Karpinski and Ø. Fischer1  Phys. Rev. Lett. \textbf{89}, 187003 (2002)
\bibitem{note}The name
LFM is inspired by the resemblance of our method  to that of a
fisherman waiting stationary with his rod for passing by fish.
\bibitem{Karpinski}J.Karpinski, M.Angst, J.Jun, S.M.Kazakov, R.Puzniak, A.Wisniewski, J.Roos, H.Keller, A. Perucchi, L. Degiorgi, M.Eskildsen, P.Bordet,
L.Vinnikov, A.Mironov Supercond. Sci. Technol. 16, 221 (2003)

\bibitem{Brinkman}A. Brinkman, A. A. Golubov, H. Rogalla, O. V. Dolgov, J. Kortus, Y. Kong, O. Jepsen, and O. K. AndersenPhys. Rev. B \textbf{65}, 180517 (2002)
\bibitem{lyard} L. Lyard, P. Szabó, T. Klein, J. Marcus, C. Marcenat, K. H. Kim, B. W. Kang, H. S. Lee, and S. I. Lee Phys. Rev. Lett. \textbf{92},
057001-1 (2004)
\bibitem{kohen2} A. Kohen, T. Cren, Th. Proslier, Y. Noat, F. Giubileo, F. Bobba,  A. M. Cucolo, W. Sacks and D. Roditchev unpublished
\end{document}